\documentclass[lettersize,journal]{IEEEtran}
\usepackage{amsmath,amsfonts}
\usepackage{algorithmic}
\usepackage{algorithm}
\usepackage{array}
\usepackage[caption=false,font=normalsize,labelfont=sf,textfont=sf]{subfig}
\usepackage{textcomp}
\usepackage{stfloats}
\usepackage{url}
\usepackage{verbatim}
\usepackage{graphicx}
\usepackage{cite}

\begin{document}

\title{Resilience-Assuring Hydrogen-Powered Microgrids}

\author{Chaofan Lin, Peng Zhang, Xiaonan Lu 

\thanks{This work relates to Department of Navy award N00014-23-1-2124 issued by the Office of Naval Research. The United States Government has a royalty-free license throughout the world in all copyrightable material contained herein.} 

\thanks{ C. Lin and P. Zhang are with the Department of Electrical and Computer Engineering, Stony Brook University, Stony Brook, NY 11794, USA (e-mail: \{chaofan.lin; p.zhang\}@stonybrook.edu).} 
\thanks{X. Lu is with the School of Engineering
Technology, Purdue University, West Lafayette, IN 47906, USA (e-mail: lu998@purdue.edu)}
}

\markboth{
}%
{Shell \MakeLowercase{\textit{et al.}}: A Sample Article Using IEEEtran.cls for IEEE Journals}


\maketitle

\begin{abstract}
Green hydrogen has shown great potential to power microgrids as a primary source, yet the operation methodology under extreme events is still an open area. To fill this gap, this letter establishes an operational optimization 
strategy towards resilient hydrogen-powered microgrids, where the frequency and voltage regulation characteristics of hydrogen sources under advanced controls are accurately represented by piecewise linear constraints. The results show that the 
new operation approach can output a safety-assured operation plan with rational power change distribution and reduced frequency and voltage variation.
\end{abstract}


\section{Introduction}
\IEEEPARstart{I}{ncreasingly} frequent natural disasters and cyber/ physical attacks pose an urgent need to build resilient microgrids in communities \cite{My_IJEPES}. Clean hydrogen and hydrogen-based fuels provide an opportunity to increase the resilience of microgrids while reducing carbon emission \cite{Hydrogen_policy}. Although there exist off-the-shelf models for optimal operation of microgrids 
\cite{Traditional_model}, the hydrogen sources have not yet been taken as the primary source to support microgrid operation. 
Further, the frequency and voltage regulation characteristics \cite{Grid-forming HS} need to be considered in the optimization so as to ensure a resilient operation of microgrids under disturbances. Thus, this letter will formulate a new scenario-based stochastic optimization with frequency and voltage regulations, and provide key insights on how primary hydrogen sources as well as the advanced control could contribute to the system resilience.

\section{Optimization Formulation}

Once the hydrogen-based sources (e.g., fuel cells or microturbines) 
are equipped with advanced droop controls, they are able to  provide frequency and voltage regulations and maintain stable operations of microgrids. For this reason, hydrogen-powered microgrids are able to maintain continuous power supply to the critical loads therein upon the occurrence of blackouts.  
To be used in extreme conditions, the optimization model in this letter assumes: (1) The microgrid is islanded; (2) The states of components, including hydrogen sources, other distributed energy resources (DERs), branches, etc., are known and only those survived are taken into account; (3) The communication system 
is down and thus the automatic secondary and tertiary control of DERs are invalid; (4) Manually adjusting the references of DERs and connections of loads is available but needs time; (5) The time-series 
curves 
for renewable sources and loads within the optimization time window can be 
forecast with some errors, while the specific forecast method is beyond this letter's focus.

The objective 
is to maximize the weighted served loads under uncertainty scenarios:
\begin{equation}\label{objective}
obj=\max{\sum_{s\in \pmb{S}}\eta_s\sum_{t\in \pmb{T}}\sum_{i\in \pmb{N}}\lambda_{i,t} w_i P^L_{s,i,t}}
\end{equation}
where $\pmb{S}$ is the set of scenarios of renewable power outputs and loads which can be generated by adding the samples from the probability distribution of forecast error to the forecast time-series power curve; $\eta_s$ is the weight of scenario $s$; $\pmb{T}$ is the set of decision moments; $\pmb{N}$ is the set of buses; $\lambda_{i,t}$ is the binary variable to decide whether to serve the load at bus $i$ or not at moment $t$; $w_i$ is the weight; $P^L_{s,i,t}$ is the active power of the load at bus $i$, moment $t$ and scenario $s$.

\emph{1) Hydrogen Constraints}

Considering the separate electrolyzer and fuel cell structure, and the active power limits, the $f-P$ characteristic for droop-controlled hydrogen sources can be formulated as:
\begin{equation}\label{HS_fP_E}
P^{He}_{s,i,t}=\left\{
\begin{aligned}
&P^{HeM}_i,f_{s,t}\geq f^{HeM}_{i,t}\\
&P^{HeM}_i-D^{HeP}_i(f^{HeM}_{i,t}-f_{s,t}),\text{others}\\
&0,P^{HeM}_i-D^{HeP}_i(f^{HeM}_{i,t}-f_{s,t})\leq 0
\end{aligned}
\right.,i\in\pmb{N}_{H}
\end{equation}
\begin{equation}\label{HS_fP_F}
P^{Hf}_{s,i,t}=\left\{
\begin{aligned}
&P^{HfM}_i,f_{s,t}\leq f^{HfM}_{i,t}\\
&P^{HfM}_i-D^{HfP}_i(f_{s,t}-f^{HfM}_{i,t}),\text{others}\\
&0,P^{HfM}_i-D^{HfP}_i(f_{s,t}-f^{HfM}_{i,t})\leq 0
\end{aligned}
\right.,i\in\pmb{N}_{H}
\end{equation}
where $\pmb{N}_{H}$ is the set of buses with a hydrogen source; $P^{He}_{s,i,t}$ and $P^{Hf}_{s,i,t}$ are the active power input of electrolyzer and output of fuel cell; $P^{HeM}_i$ and $P^{HfM}_i$ are the maximum input of electrolyzer and output of fuel cell; $f^{HeM}_{i,t}$ and $f^{HfM}_{i,t}$ are the maximum frequency of droop mode for electrolyzer and minimum frequency of droop mode for full cell; $D^{HefP}_i$ and $D^{HffP}_i$ are the corresponding droop coefficients.

Noted that the piecewise equations (\ref{HS_fP_E}) and (\ref{HS_fP_F}) under droop control can be used for both grid-forming and grid-following inverters, but (\ref{HS_fP_E}) and (\ref{HS_fP_F}) are usually reversely formulated as $P-f$ functions in a grid-forming mode.

To choose either electrolyzer or fuel cell to work, we have:
\begin{equation}\label{HS_choose1}
P^{H}_{s,i,t}=x^{Hf}_{i,t}P^{Hf}_{s,i,t}-x^{He}_{i,t}P^{He}_{s,i,t},i \in \pmb{N}_{H}
\end{equation}
\begin{equation}\label{HS_choose2}
x^{He}_{i,t}+x^{Hf}_{i,t}\leq 1,i \in \pmb{N}_{H}
\end{equation}
where $P^{H}_{s,i,t}$ is the active power injection of hydrogen; $x^{He}_{i,t}$ and $x^{Hf}_{i,t}$ are the binary variables to decide whether the electrolyzer or the full cell is working.

Also, the hydrogen tank constraints should be considered:
\begin{equation}\label{HS_tank1}
H_{s,i,t}=H_{s,i,t-1}-x^{Hf}_{i,t}P^{Hf}_{s,i,t}/\eta^{Hf}_i+x^{He}_{i,t}P^{He}_{s,i,t}\eta^{He}_i, i \in \pmb{N}_{H}
\end{equation}
\begin{equation}\label{HS_tank2}
0\leq H_{s,i,t}\leq H^{max}_i, i \in \pmb{N}_{H}
\end{equation}
where $H_{s,i,t}$ is the amount of equivalent hydrogen energy; $H^{max}_i$ is the maximum hydrogen storage; $\eta^{He}_i$ and $\eta^{Hf}_i$ are the efficiency rates of electricity-to-hydrogen and hydrogen-to-electricity conversion for the electrolyzer and fuel cell.

For voltage regulation, the hydrogen sources can also provide droop support within their reactive power output limits \cite{Grid-forming HS}. A dead band can be set to avoid frequent switching between reactive power generation and absorption modes \cite{Vol-var standard}:
\begin{equation}\label{RES_HS_UQ}
Q^{H}_{s,i,t}=\left\{
\begin{aligned}
&Q^{GM}_i,D^{GQ}_i(U^{GM}_{i,t}-U_{s,i,t})>Q^{GM}_i\\
&D^{GQ}_i(U^{GM}_{i,t}-U_{s,i,t}),U_{s,i,t}\leq U^{GM}_{i,t}\ \&\\
&\quad\quad\quad D^{GQ}_i(U^{GM}_{i,t}-U_{s,i,t})\leq Q^{GM}_i\\
&0,U^{GM}_{i,t}<U_{s,i,t}<U^{AM}_{i,t}\quad\quad\quad\quad,i\in\pmb{N}_{H}\\
&-D^{AQ}_i(U_{s,i,t}-U^{AM}_{i,t}),U_{s,i,t}\geq U^{AM}_{i,t}\ \&\\
&\quad\quad\quad D^{AQ}_i(U_{s,i,t}-U^{AM}_{i,t})\leq Q^{AM}_i\\
&-Q^{AM}_i,D^{AQ}_i(U_{s,i,t}-U^{AM}_{i,t})>Q^{AM}_i
\end{aligned}
\right.
\end{equation}
where $Q^{H}_{s,i,t}$ is the reactive power output of hydrogen; $Q^{GM}_i$ and $Q^{AM}_i$ are the maximum generated and absorbed reactive power; $U^{GM}_{i}$ and $U^{AM}_{i}$ are the boundaries of voltage dead band to start reactive power generation and absorption; $D^{GUQ}_i$ and $D^{AUQ}_i$ are the corresponding droop coefficients.

\emph{2) Other DERs' Constraints}

This letter only presents the constraints of renewable sources as the representation for other DERs, since they are usually co-operated with hydrogen sources in microgrids \cite{Traditional_model}. Considering the environment-dependent maximum power point (MPP), a renewable source has the following $f-P$ characteristic \cite{Grid-forming RES}:
\begin{equation}\label{RES_fP}
P^{R}_{s,i,t}=\left\{
\begin{aligned}
&\tilde{P}^{RM}_{s,i,t},f_{s,t}\leq f^{RM}_{i,t}\\
&\tilde{P}^{RM}_{s,i,t}-D^{RP}_i(f_{s,t}-f^{RM}_{i,t}),\text{others}\\
&0,\tilde{P}^{RM}_{s,i,t}-D^{RP}_i(f_{s,t}-f^{RM}_{i,t})\leq 0
\end{aligned}
\right.,i\in\pmb{N}_{R}
\end{equation}
where $\pmb{N}_{R}$ is the set of buses with a renewable source; $P^{R}_{s,i,t}$ and $\tilde{P}^{RM}_{s,i,t}$ are the actual active power output and uncertain MPP; $f_{s,t}$ is the system frequency; $f^{RM}_{i,t}$ is the minimum frequency of droop mode; $D^{RP}_i$ is the droop coefficient.

The voltage regulation characteristic of renewable sources is also (\ref{RES_HS_UQ}) and the reactive power output is denoted as $Q^{R}_{s,i,t}$.

\emph{3) Network Constraints}

The KCL constraints can then be formulated as below:
\begin{equation}\label{KCL1}
\sum_{ik\in\pmb{B}}P_{s,ik,t}=\sum_{ji\in\pmb{B}}P_{s,ji,t}+P^R_{s,i,t}+P^H_{s,i,t}-\lambda_{i,t}P^L_{s,i,t}, i \in \pmb{N}
\end{equation}
\begin{equation}\label{KCL2}
\sum_{ik\in\pmb{B}}Q_{s,ik,t}=\sum_{ji\in\pmb{B}}Q_{s,ji,t}+Q^R_{s,i,t}+Q^H_{s,i,t}-\lambda_{i,t}Q^L_{s,i,t}, i \in \pmb{N}
\end{equation}
where $\pmb{B}$ is the set of all branches; $P_{s,ik,t}$ and $Q_{s,ik,t}$ are the active and reactive power of the branch from bus $i$ to bus $k$.

The KVL constraints by linearized DistFlow model are:
\begin{equation}\label{KVL}
U_{s,j,t}-U_{s,i,t}=(R_{ij}P_{s,ij,t}+X_{ij}Q_{s,ij,t})/U_0, ij\in\pmb{B}
\end{equation}
where $R_{ij}$ and $X_{ij}$ are the resistance and reactance of the branch; $U_0$ is the nominal voltage.

\emph{4) Security Constraints}

To ensure microgrid's security, the system frequency, bus voltage and branch flow should be limited within their ranges:
\begin{equation}\label{security1}
U^{min}_i\leq U_{s,i,t}\leq U^{max}_i, i \in \pmb{N}
\end{equation}
\begin{equation}\label{security2}
f^{min}\leq f_{s,t}\leq f^{max}
\end{equation}
\begin{equation}\label{security3}
P_{ij}^{min}\leq P_{s,ij,t}\leq P_{ij}^{max}, ij\in\pmb{B}
\end{equation}
\begin{equation}\label{security4}
Q_{ij}^{min}\leq Q_{s,ij,t}\leq Q_{ij}^{max}, ij\in\pmb{B}
\end{equation}
where $U^{min}_i$ and $U^{max}_i$ are the minimum and maximum bus voltage; $f^{min}$ and $f^{max}$ are the minimum and maximum system frequency; $P_{ij}^{min}$, $P_{ij}^{max}$, $Q_{ij}^{min}$ and $Q_{ij}^{max}$ are the minimum and maximum active and reactive branch flow.

The above optimization model has several piecewise linear constraints like (\ref{HS_fP_E}) and nonlinear product terms like (\ref{HS_choose1}) which can be easily converted to linear constraints by using the big-M method. Therefore, the model after conversion belongs to the mixed integer linear programming (MILP) and can be solved with commercial solvers like Gurobi, CPLEX, etc.

\section{Case Study}
The proposed optimization model is validated on a modified balanced IEEE 13 bus test feeder, where Bus 645 is integrated with a hydrogen source to power the microgrid, and Bus 633 and 680 are integrated with two wind farms. The load weight is randomly selected from $(0,1)$ and those loads with a weight larger than 0.7 are regarded as critical loads. The optimization time window is set to 6 hours, and the 15-min interval forecast data for wind farms and electric loads are from the ENTSO-E Transparency Platform and are scaled down to fit the magnitude of this microgrid. The forecast error is assumed to follow the normal distribution $N(0,10\%P_{max})$. Two extreme scenarios with maximum and minimum average power and the original forecast scenario are used for the optimization. Their probabilities are set to $0.001$, $0.001$ and $0.998$ so to maintain the economy and robustness of the optimized strategy.

\emph{1) Model Validation}

\begin{figure}[t]
\centering
\includegraphics[width=3.4in]{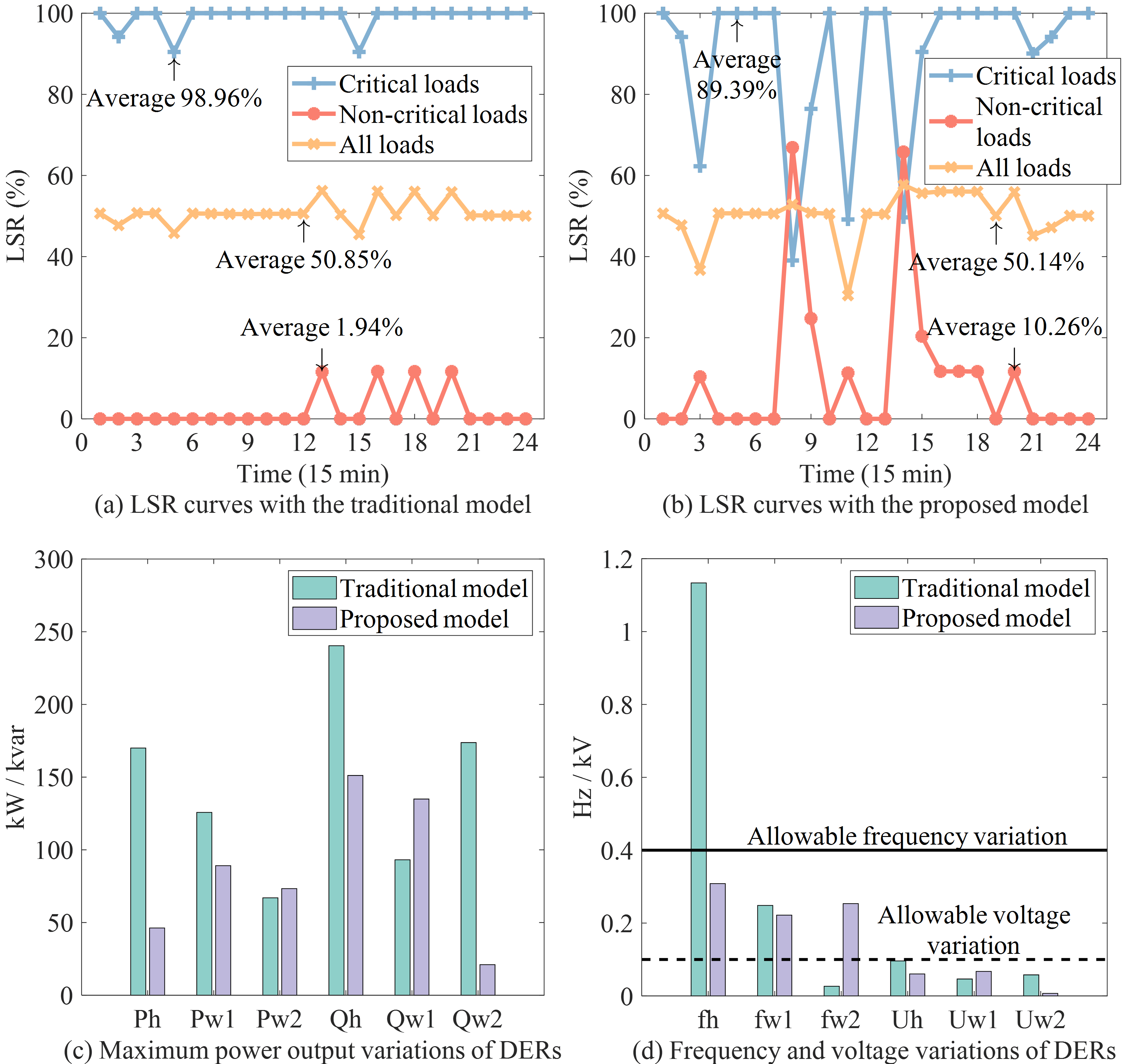}
\caption{LSR, power outputs, and frequency/voltage variation comparison.}
\label{fig_1}
\end{figure}

The proposed optimization model with frequency and voltage regulations is compared with the state-of-the-art model in \cite{Traditional_model}. The load served ratio (LSR) indexes of the critical loads, non-critical loads and all loads during the time window using the two models are shown in Fig. \ref{fig_1}(a) and (b). The maximum power output variations and corresponding frequency and voltage variations of DERs are shown in Fig. \ref{fig_1}(c) and (d).

Comparing Fig. \ref{fig_1}(a) and (b), it can be 
seen that with the proposed model, at some operation periods, the LSR of the critical loads decreases while the LSR of the non-critical loads increases, which brings the average values from 98.96\% down to 89.39\% and 1.94\% up to 10.26\% respectively. The average LSR of all loads is not significantly affected. However, despite of the seemingly unsatisfactory change in LSR, the proposed model can dramatically reduce the power output variation and thus the frequency and voltage variation as well, as can be seen from Fig. \ref{fig_1}(c) and (d). This is because the distribution of sudden uncertain power change in the traditional model is irrational, or is decided by other economic factors, ignoring the effect of droop control on power distribution and differing from real system configuration, which could easily violate the frequency or voltage limitations and cause stability problems.

\emph{2) Resilience Contribution of Hydrogen Sources}

\begin{figure}[t]
\centering
\includegraphics[width=3.4in]{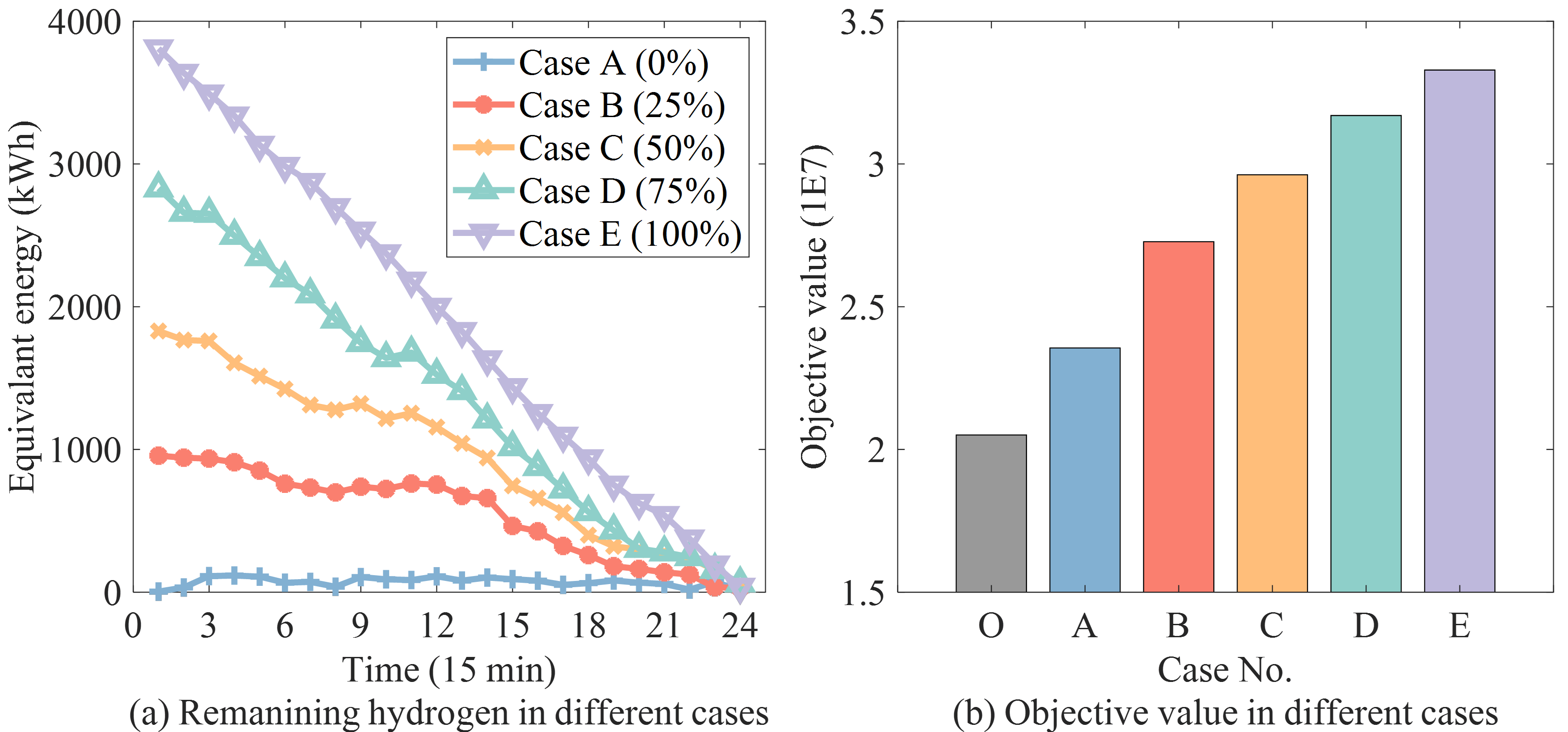}
\caption{Optimization results of cases with different initial stored hydrogen.}
\label{fig_2}
\end{figure}

To quantify the contribution of hydrogen sources to system resilience, five cases numbered A-E are created with 0 to 100\% initial stored hydrogen with a 25\% increment. Case O is also created where the hydrogen source is removed. The states of average remaining hydrogen under different scenarios and the objective values are shown in Fig. \ref{fig_2}(a) and (b).

It can be seen from Fig. \ref{fig_2}(a) that, with the proposed optimization model, the hydrogen source can actively participate in the energy management even without initial stored hydrogen. More importantly, Fig. \ref{fig_2}(b) indicates that installing a hydrogen source can greatly improve the objective value (i.e., system resilience) with or without initial stored hydrogen. Besides, with increased initial stored hydrogen, the increment of its contributed resilience 
decreases, which means the hydrogen source might have the most marginal contribution to system resilience at a moderate initial stored hydrogen amount.

\emph{3) Resilience Contribution of Advanced Droop Control}

To quantify the contribution of advanced droop control to system resilience, three cases numbered I-III are created with one up to three DERs with frequency and voltage regulations. The hydrogen source, as the primary source in hydrogen-powered microgrids, provides droop regulation service in all the three cases. The renewable sources can work in either constant $PQ$ mode or droop mode. Table \ref{tab_1} shows some operation indexes.
\begin{table}[t]
\caption{Operation Indexes in Different Numbers of Grid-Forming DERs\label{tab_1}}
\centering
\begin{tabular}{c|c c c}
\hline
Average Indexes & Case I & Case II & Case III\\
\hline
Objective Value (1E7) & 2.3764 & 2.8527 & 3.1698\\
Average LSR of All Loads (\%) & 39.68 & 47.35 & 50.14\\
Average LSR of Critical Loads (\%) & 64.90 & 75.19 & 89.39\\
Renewable Consumption Ratio (\%) & 69.76 & 89.13 & 95.71\\
Average Frequency Variation (Hz) & 0.1605 & 0.1826 & 0.1646\\
Average Voltage Variation (V) & 6.5032 & 5.5166 & 5.5231\\
\hline
\end{tabular}
\end{table}

It can be learned from Table \ref{tab_1} that, with increased number of DERs with frequency and voltage regulations, the objective value, the LSR of all loads, the LSR of critical loads and the renewable consumption ratio are all improved significantly, by 33.39\%, 26.36\%, 37.73\% and 37.20\% respectively. The LSR of critical loads has a most dramatic increase. Meanwhile, the frequency and voltage variation shows both increasing and decreasing trends. This is because they are affected by both the uncertainty and regulation levels. More uncertainty from renewable and served loads will amplify the variation but more DERs participating in regulation will conversely suppress it.

\section{Conclusion}

This letter 
demonstrate how to incorporate the frequency and voltage regulation characteristics of 
hydrogen sources and other DERs 
into the optimization of hydrogen-powered microgrid operation. The results show that the proposed optimization model can more rationally distribute the uncertain powers 
and greatly reduce the frequency and voltage variations, providing a more realistic and resilient operation strategy for microgrid operators to counteract extreme events and uncertainties. 
In the future, the detailed characteristics of hydrogen sources will be modeled to improve the practicability.

\vfill

\end{document}